\documentclass[aps,pre,twocolumn,groupedaddress,amsmath,showpacs]{revtex4}

\usepackage{graphicx}
\usepackage{psfrag}
\newcommand{\nm}{\,\mathrm{nm}}

\psfrag{rhol}[B1][B1][2.5]{$\rho_l$}
\psfrag{rhor}[B1][B1][2.5]{$\rho_r$}
\psfrag{Ee}[B1][B1][2.5]{$a=q/E$}
\psfrag{Ll}[B1][B1][2.5]{$L$}
\psfrag{jj}[B1][B1][1.0]{$\frac{j}{\rho_0}\sqrt{\frac{2\pi m}{k_BT}}$}
\psfrag{maL}[B1][B1][1.0]{$maL/k_BT$}
\psfrag{jjj}[B1][B1][1.0]{$\sqrt{\frac{m}{k_BT}}j$}
\psfrag{alp}[B1][B1][1.0]{$\alpha$}

\begin{document}


\title{Kinetic models of ion transport through a nanopore}


\author{Jaroslaw Piasecki}
\affiliation{Institute of Theoretical Physics, University of Warsaw, Ho\.{z}a 69, 00 681 Warsaw, Poland}

\author{Rosalind J. Allen}
\email{R.Allen@amolf.nl}
\altaffiliation[Current address: ]{FOM-Institute for Atomic and Molecular Physics, Kruislaan 407, 1098 SJ Amsterdam, The Netherlands}

\author{Jean-Pierre Hansen}%
\affiliation{The University Chemical Laboratory, Lensfield Road, Cambridge CB2 1EW, United Kingdom}

\date{\today}

\begin{abstract}
Kinetic equations for the stationary state distribution function of ions
moving
through narrow pores are solved for a number of one-dimensional models of
single ion transport. Ions move through pores of length $L$, under the action of a constant external field
and of a concentration gradient. The interaction of
single ions with the confining pore surface and with water molecules inside the
pore are modelled by a Fokker-Planck term in the kinetic equation, or by uncorrelated collisions with thermalizing centres distributed along the pore. The
temporary binding of ions to polar residues lining the pore is modelled by
stopping traps or energy barriers. Analytic expressions for the stationary ion
current through the pore are derived for several versions  of
the model, as functions of key physical parameters. In all cases, saturation of the current
at high fields is predicted. Such simple models, for which results are analytic,  may prove useful in the study of the 
current/voltage relations of ion channels through membranes.
\end{abstract}


\maketitle

\section{Introduction}
The flow of fluids through porous media is a classic problem which has many scientific and industrial applications. For very narrow pores, with diameters of the order of $1\nm$ or less, continuum descriptions become inapplicable and the transport of matter must be examined on the molecular scale. Examples include molecular or ionic permeation of zeolites \cite{dubbeldam}, of carbon nanotubes \cite{kalra,berezhkovskii,hummer} and of aquaporins \cite{fujiyoshi} and ion channels \cite{hille} through cell membranes. The simple kinetic models examined in this work are meant to represent ion channels: they are, however, also more widely applicable - for example, we shall present results for ions flowing through an infinitely long pore that might represent a carbon nanotube or part of a zeolite.

Ion channels are pores through cell membranes, through which ions are transported under the influence of a concentration gradient and a large electric field. The permeability of the pores is highly selective for particular ions and the pores can also open and close to ion transport (a phenomenon known as ``gating'') in response to factors such as ligand binding or changes in the electric field or the membrane tension. Many channels contain a narrow region, the ``selectivity filter'', where ionic motion is  essentially single-file \cite{doyle,morais,zhou,allen}. Some channels appear to transport only one ion at a time, while others use transport mechanisms involving multiple ions \cite{morais,zhou,berneche,berneche2} . Measurements of the current through individual channels have been possible for some time, and these have resulted in a large amount of data, both on the gating characteristics of channels, and on their properties in the open state. These properties include the relationship between the ionic current and the electric field applied across the membrane (current-voltage relations), as well as the conductivity of the channels as a function of the ionic concentration difference, at fixed applied voltage (conductance-concentration relations). One of the challenges for theoreticians is to relate these functional characteristics to the geometric, physical and chemical structure of the pores, which are becoming increasingly well-known \cite{levitt}.  This goal may be achieved by detailed simulations of the motion of ions and molecules through specific pores \cite{tieleman,kuyucak}, or using simplified models \cite{crozier,chou,chou2,chou3}. Rates of ion transport can be predicted directly, or by application of barrier-crossing theories such as Kramers rate theory \cite{hanggi,laio,tolokh}. An alternative approach is the extension of continuum theories to the nanoscale.  Goldman \cite{goldman} and Hodgkin and Katz \cite{hodgkin}(GHK), in their classic work, applied the one-dimensional diffusion equation in a constant electric field to predict current-voltage relations for ion channels. This work is generalized to specific and multi-dimensional ion channel models in the Poisson-Nernst-Planck (PNP) theory of ion channels \cite{chen,graf}, where numerical solution methods are used to obtain the current due to diffusion in the presence of complicated and self-consistent potential fields. 

In this paper, we explore the possibility of applying simple, analytically solvable, kinetic models to the problem of transport in nanopores. Our approach is for the moment very general. We consider the motion of  single ions through a one-dimensional pore of length $L$ connecting two reservoirs at different ion concentrations, under the action of a constant electric field (we shall also consider the case of an infinitely long pore). Ions are expected to experience friction due to collisions with the inside surface of the pore and with other particles, such as water molecules,  inside the pore.   We include this effect initially by a  Langevin-like friction within the framework of a Fokker-Planck (FP) description, although we shall see that the FP approach presents some difficulties in the confined geometry of the finite length pore, which we shall attempt to overcome by using an alternative description of the friction in terms of localized ``thermalizing centres''. We also consider that there may be more specific binding interactions between the ion and the pore surface, for example with polar residues lining the surface. These are modelled by ``stopping traps'' - on encountering such a trap, an ions is stopped, and later released to continue its motion under the influence of friction and the electric field. The action of the stopping traps on the ion may or may not depend upon its velocity. In all cases, we attempt to find general  analytic solutions for  the  stationary state ionic current $j$, and sometimes also for the ion distribution function $f(x,v)$ (defined so that the probability of finding an ion between position $x$ and $x+dx$ with velocity between $v$ and $v+dv$ is $f(x,v) dx dv$). These solutions are functions of the applied field, as well as of parameters such as the channel length, friction coefficient and probability density of stopping traps. We hope that these results may ultimately be used to analyse the transport behaviour of specific pores or channels, by adapting the above physical parameters to the known structure of the pore under consideration. 

The general kinetic equation for the model is presented in section \ref{sec2}. The kinetic equation is solved in section \ref{sec3} for the case of a finite length channel with stopping traps but without friction, and in section \ref{sec4} in the case of a finite channel with friction but without traps; difficulties arising from the use of the FP operator in a pore of finite length are discussed. These difficulties do not arise in the case of an infinite ($L \to \infty$) channel, for which a general solution is obtained in the presence of both friction and traps, in section \ref{sec5}. Returning to a finite $L$ channel in section \ref{sec6}, the FP friction is replaced by a distribution of thermalizing centres throughout the pore, and an explicit expression for the current is obtained in the presence of such thermalizing centres and stopping traps. Concluding remarks are made in the last section.

\section{Model and kinetic equation}\label{sec2}
\begin{figure}
\scalebox{0.4}{\rotatebox{0}{\includegraphics{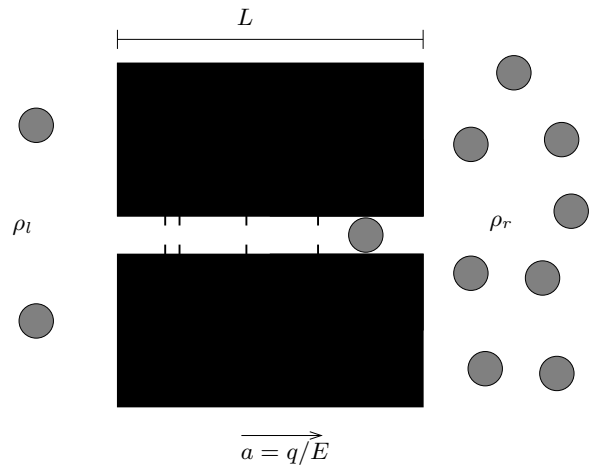}}}
\caption{Schematic view of the model channel.\label{fig1}}
\end{figure}

Our model channel is pictured in figure \ref{fig1}. The channel, of length $L$, is located along the  $x$-axis $(-L/2 < x < L/2)$. The radius of the pore (which is assumed to be cylindrical) matches the ion radius, so that ionic motion inside the pore is strictly one-dimensional. The pore links two reservoirs containing ionic solutions of linear concentrations $\rho_l$ (to the left) and $\rho_r$ (to the right): $\rho_l$ and $\rho_r$ are related to the bulk concentrations $c_l$ and $c_r$ in the reservoirs by $\rho_{r,l} = \pi R^2 c_{r,l} $, where $R$ is the radius of the pore.  The inner surface of the pore is lined with stopping traps of local average density $\rho(x)$: we shall assume that the probability of finding $n$ traps within the interval $x_1 < x < x_2$ is given by the Poisson distribution with parameter $\int_{x_1}^{x_2}\rho(x)dx$.

 If an ion encounters a trap, its velocity is set to zero, generally irrespective of its initial velocity (although we shall also consider in section \ref{sec3} the case of traps which discriminate between ions according to velocity). Inside the pore, the ion (of charge $q$ and mass $m$) is subjected to a uniform electric field $E$, and hence undergoes an acceleration, towards the right, $a=qE/m$. After being stopped by a trap, the ion is therefore re-accelerated by the electric field. Ions also experience friction: this will initially be modelled by a force $-\gamma v$, where $v$ is the velocity of the ion, as well as the thermalizing effect of a random force, although an alternative to this Langevin-like  model will be presented in section \ref{sec6}. In summary, the ion undergoes a constant acceleration due to the electric field, is slowed down by collisions with molecules inside the pore or on the pore surface (these processes being described by a friction process) and may be captured by traps along the channel, to account for temporary binding to polar residues on the pore surface. We shall present results for stationary state ion flow only.

The general kinetic equation for the stationary state ion distribution function $f(x,v)$, in the presence of Poisson-distributed stopping traps of average density $\rho(x)$ as well as a Langevin-like friction mechanism, with friction coefficient $\gamma$, is:
 \begin{eqnarray}\label{eq1}
&& \left(v\frac{\partial}{\partial x} + a\frac{\partial}{\partial v}\right)f(x,v) \\*\nonumber &&= \rho(x) \left\{ \delta(v)\int_{-\infty}^{\infty} dw\, |w|f(x,w) - |v| f(x,v)\right\}  \\*\nonumber && \,\, + \gamma\frac{\partial}{\partial v}\left(v+\frac{k_BT}{m}\frac{\partial}{\partial v}\right) f(x,v)
\end{eqnarray}
The left-hand side (l.h.s.) of (\ref{eq1}) describes free flow of ions under the action of the constant acceleration $a$ arising from the external field. The r.h.s. contains two collision terms. The first accounts for the stopping traps: it is a balance between gain (in the population of zero velocity particles) and loss (of particles with velocity $v$). The second term is the Fokker-Planck operator acting on the distribution function: it accounts for the effect of the frictional and random forces. Note that the kinetic equation (\ref{eq1}) is for a single ion: it does not account for interactions between several ions within the pore. This limitation will be addressed in later work.

The reservoirs on the left ($x < -L/2$) and on the right ($x>L/2$) of the channel are assumed to contain ions in thermodynamic equilibrium at the same temperature $T$, but generally at different densities: $\rho_l$ (to the left) and $\rho_r$ (to the right). The ion distribution functions in the reservoirs (not including the contribution of any ions coming out of the pore) are hence:
\begin{equation}\label{eq2}
f_l(x,v) = \rho_l\phi^T(v) \qquad  f_r(x,v) = \rho_r\phi^T(v)
\end{equation}
where 
\begin{equation}\label{eq3}
\phi^T(v) = \sqrt{\frac{m}{2 \pi k_BT}}\exp{\left[-\frac{mv^2}{2k_BT}\right]}
\end{equation}
is the Maxwell velocity distribution function.

For illustrative purposes, we first consider the case where acceleration, traps and friction are all absent, and an ion which enters the pore at one end keeps the same velocity until it reaches the other end. The ion distribution function within the pore is then simply:
\begin{equation}\label{eq4}
f(x,v) = \left[\rho_l \theta(v) + \rho_r \theta(-v)\right]\phi^T(v)
\end{equation}
where $\theta$ denotes the Heaviside step function. The ion current is given by:
\begin{equation}\label{eq5}
j(x) = \int_{-\infty}^{\infty} dv\, v f(x,v)
\end{equation}
For a stationary state, continuity requires that the current be independent of position:
\begin{equation}\label{eq6}
\frac{d j(x)}{dx} = 0
\end{equation}
Substituting (\ref{eq4}) into (\ref{eq5}) one finds the result in the absence of acceleration, traps or friction:
\begin{equation}\label{eq7}
j=\sqrt{\frac{k_BT}{2 \pi m}}(\rho_l-\rho_r)
\end{equation}
while the number density inside the channel
\begin{equation}\label{eq8}
n(x) = \int_{-\infty}^{\infty} dv\, f(x,v)
\end{equation} 
is in this case given by $n=(\rho_l+\rho_r)/2$. Note that discontinuities arise in $n(x)$ at the pore boundaries, $x=\pm L/2$: this reflects the fact that the regions close to the pore mouth are not modelled in detail in this simple theory.

In the subsequent sections, analytic solutions of the kinetic equation (\ref{eq1}) will be derived for the limiting cases $\gamma=0$ (section \ref{sec3}), $\rho(x)=0$ (section \ref{sec4}) and $L\to \infty$ (section \ref{sec5}).


%



\section{Finite channel with traps}\label{sec3}
Consider a pore of finite length $L$, containing stopping traps but no friction mechanism. The traps have average local density $\rho(x)$ and are distributed according to a Poisson law as described in Section \ref{sec2}: on encountering such a trap, the velocity of an ion is reduced to zero.   In the absence of friction,  the kinetic equation (\ref{eq1}) for the stationary state ion distribution function $f(x,v)$ simplifies to:
\begin{eqnarray}\label{eq9}
&&\left(v\frac{\partial}{\partial x} + a \frac{\partial}{\partial v} \right) f(x,v) \\*\nonumber && = \rho(x) \left\{\delta(v) \int_{-\infty}^{+\infty}dw |w| f(x,w)-|v|f(x,v)\right\}
\end{eqnarray}
 Note that we obtain Equation (\ref{eq6}) (constant current throughout the pore) on integrating both sides of (\ref{eq9}) over all velocities $-\infty < v < +\infty$. If the traps are on average uniformly distributed ($\rho(x) = \rho$), Equation (\ref{eq9}) can be solved exactly for $f(x,v)$ as shown in Appendix \ref{appa}. However, an expression for the ionic current $j$ can be obtained for any $\rho(x)$ using simple arguments,  without the need for an explicit solution for $f(x,v)$.

We first note that in the stationary state, the contribution to the current due to an ion which enters the channel at one end depends only on its incoming velocity and its probability of eventually arriving at the other end, since $j$ does not depend on $x$ (and there are no interactions between ions). For this model, all ions entering the channel from the left reservoir at $x=-L/2$ will eventually reach $x=L/2$, since on being stopped by a trap they are re-accelerated by the field towards the right (assuming $a$ is positive). Thus the contribution of these ions to the current is: 
\begin{equation}\label{eq11}
j_l=\rho_l\sqrt{\frac{k_BT}{2\pi m}}
\end{equation} 
Ions entering the channel from the right at $x=L/2$ will reach $x=-L/2$ only if they are not stopped either by the opposing field or by an encouter with a trap. An ion which is stopped is re-accelerated towards the right, so that it will exit the channel at $L/2$. The Poisson probability of encountering no traps between $x_1$ and $x_2$ is:
\begin{equation}\label{eq12}
P(x_1,x_2) = \exp{\left\{-\int_{x_1}^{x_2} \rho(x') dx'\right\}}\qquad  x_1 < x_2
\end{equation} 
In order to overcome the opposing field, ions must enter the channel with kinetic energy $m v^2/2 > maL$, so that, assuming a Maxwell distribution of velocities at $L/2$, the distribution function $f_r(x,v)$ of particles which entered the channel at $L/2$ and which will eventually reach $-L/2$ is:
\begin{eqnarray}\label{eq13}
f_r(x,v) =&& \rho_r \theta(-v)\, \theta \left(\frac{v^2}{2} + a\left(\frac{L}{2}-x\right)-aL\right)\\*\nonumber && \times \sqrt{\frac{m}{2 \pi k_BT}}\exp{\left\{-\frac{m}{k_BT}\left[\frac{v^2}{2} + a\left(\frac{L}{2}-x\right)\right]\right\}}\\*\nonumber && \times P\left(-\frac{L}{2},\frac{L}{2}\right)
\end{eqnarray}
from which Equation (\ref{eq5}) leads to a contribution to the current:
\begin{equation}\label{eq14}
j_r = -\rho_r \sqrt{\frac{k_BT}{2 \pi m}} \exp{\left\{-\frac{maL}{k_BT}-\int_{-L/2}^{L/2}\rho(x') dx'\right\}}
\end{equation}
Adding (\ref{eq11}) and (\ref{eq14}) leads to the result for $j$:
\begin{equation}\label{eq15}
j = \sqrt{\frac{k_BT}{2 \pi m}} \left[ \rho_l -\rho_r \exp{\left\{-\frac{maL}{k_BT}-\int_{-L/2}^{L/2}\rho(x') dx'\right\}}\right]
\end{equation}
For a uniform distribution of traps, Equation (\ref{eq15}) reduces to:
\begin{equation}\label{eq10}
j=\sqrt{\frac{k_BT}{2\pi m}}\left[\rho_l-\rho_r\exp{\left\{-L\left(\rho+\frac{ma}{k_BT}\right)\right\}}\right]
\end{equation}
in agreement with the full solution derived in Appendix \ref{appa}. Equation (\ref{eq15}) shows that in this model, in the absence of friction, the current does not depend on the spatial distribution of the stopping traps, but only on the integral of $\rho(x)$ between $x=-L/2$ and $x=L/2$. Note that Equations (\ref{eq15}) and (\ref{eq10})  were derived for $a>0$. When $a<0$, the roles of the right and left reservoirs must be interchanged.

\begin{figure}
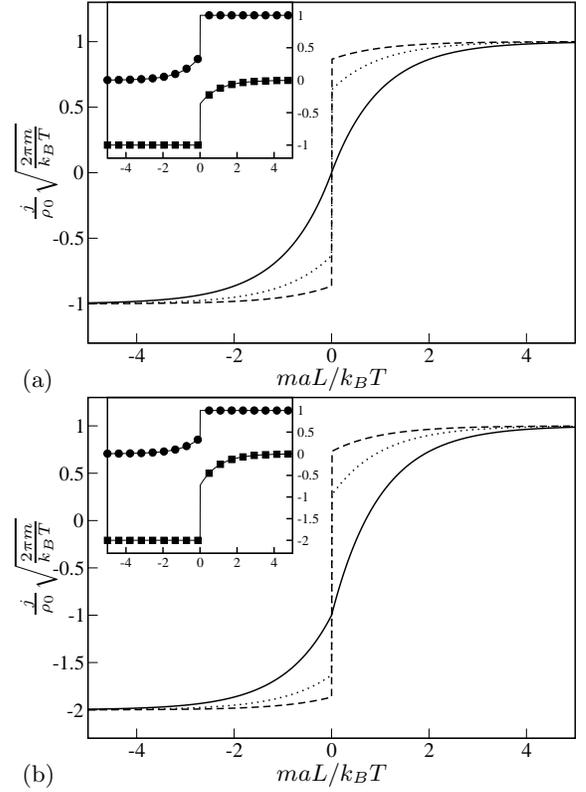

\begin{center}
\makebox[0pt][l]{(a)}{\rotatebox{0}{{\includegraphics[scale=0.3,clip=true]{fig2a.eps}}}}\hspace{0.5cm}\makebox[0pt][l]{(b)}{\rotatebox{0}{{\includegraphics[scale=0.3,clip=true]{fig2b.eps}}}}
\caption{Dimensionless current $\sqrt{2\pi m/(k_BT)}j/\rho_0$ as a function of $maL/(k_BT)$, where the reservoir densities are $\rho_l=C_l \rho_0$ and $\rho_r=C_r \rho_0$, for various values of the dimensionless stopping trap density $\rho L$. Solid lines: $\rho L=0$, dotted lines: $\rho L=1$, dashed lines: $\rho L=2$. (a): Equal reservoir densities, $\rho_r=\rho_l$; $C_r=1$, $C_l=1$ (b): $\rho_r=2\rho_l$; $C_r=2$, $C_l=1$. The insets show the currents $j_l$ and $j_r$ (in dimensionless form) due to ions originating in the left (circles) and right (squares) reservoirs, for the case $\rho L=1$. Values for the current in absolute units can be obtained by substituting absolute values for the physical parameters $a$, $m$, $L$, $\rho_r$, $\rho_l$, $\rho$ and $k_BT$.\label{fig2}}
\end{center}
\end{figure}

If the reservoir densities $\rho_l$ and $\rho_r$ are measured  relative to an arbitrary ``reference density'' $\rho_0$, such that $\rho_l=C_l \rho_0$ and $\rho_r=C_r \rho_0$, a dimensionless form of the current is given by $\sqrt{2\pi m/(k_BT)}j/\rho_0$.  This is plotted in Figure \ref{fig2}, for the cases where the reservoir densities are equal ($C_l=C_r$) or different ($C_l < C_r$). Figure \ref{fig2} shows that the current saturates for large $|a|$.  For positive $a$, the current at saturation is due exclusively to ions from the left reservoir and $\sqrt{2\pi m/(k_BT)}j/\rho_0 \to C_l$; for negative $a$, $\sqrt{2\pi m/(k_BT)}j/\rho_0  \to C_r$.  For values of $|a|$ below saturation, the magnitude of the current increases as the density of traps increases. This is because traps reduce the negative current contribution $j_r$ from the right reservoir (for $a>0$), without affecting the current $j_l$ of ions moving from the left, as can be seen in the insets, where $j_r$ and $j_l$ (in dimensionless form) are plotted individually for the case where $\rho L=1$. The discontinuity in the current at $a=0$, observed for finite concentrations of traps ($\rho > 0$), reflects the fact that the model is no longer valid in the absence of a field, when there is no stationary solution (since ions that are stopped by a trap are not then re-accelerated).
In the case of unequal ion densities in the two reservoirs, the current-voltage curves are asymmetric, as shown in Figure \ref{fig2}b. The saturation value of $|j|$ is now larger for negative $a$, and $j$ is negative for small positive values of $a$.

Thus far, we have assumed that any ion which encounters a trap is stopped, regardless of its velocity. However, ions with low kinetic energy could be expected to be more likely to be bound by a polar residue lining a nanopore, than those with more energy. We now consider a variation on our previous model, in which a single trap is present at position $x=x_0$ inside the pore ($-L/2 < x_0 < L/2$), which presents an ``energy barrier'' of height $E_0=mv_0^2/2$ to all ions crossing $x=x_0$. We shall consider two possible modes of action of this trap.

In model A, the trap at $x=x_0$ stops all ions  with kinetic energy below the barrier height: $m v^2/2 < E_0$ (and subsequently releases them to be  re-accelerated by the electric field), but has no effect on ions with energy  $mv^2/2 > E_0$. The appropriate kinetic equation reads:
\begin{eqnarray}\label{eq16}
&&\left(v\frac{\partial}{\partial x} + a \frac{\partial}{\partial v} \right) f(x,v) = \delta(x-x_0) \times \\*\nonumber && \left\{\delta(v)\int_{-v_0}^{v_0}dw |w| f(x,w)-\theta(v_0-|v|)|v|f(x,v)\right\}
\end{eqnarray}
Equation (\ref{eq16}) can be solved analytically, but we shall instead use simple arguments, as before, to obtain the current $j$ without the explicit form of $f(x,v)$. As above, any ion entering the channel at the left extremity ($x=-L/2$) with velocity $v>0$ will eventually reach the right extremity ($x=L/2$), regardless of whether it is stopped by the trap. The contribution $j_l$ of these ions to the current is therefore given by (\ref{eq11}). However, ions  entering the channel at $x=L/2$ with velocity $v<0$ will only reach $x=-L/2$ (and hence contribute to the current) if they are not stopped either by the field or by the energetic trap at $x_0$. There are two possibilities, depending on the barrier height $E_0$:

(1) $E_0 > ma \left(x_0 + L/2\right)$. In this case, any ion which reaches the trap with energy greater than $E_0$, and so is not stopped, must have  sufficient energy to overcome the remaining part of the opposing field between $x_0$ and $-L/2$.  In order to have energy $\ge E_0$ on reaching the trap at $x_0$, an ion must enter the channel at $x=L/2$ with velocity $v$ such that:
\begin{equation}\label{eq17}
\frac{mv^2}{2} > E_0 + ma \left(\frac{L}{2}-x_0 \right)
\end{equation}
so that, assuming an incoming Maxwell distribution, the distribution function of these ions is:
\begin{eqnarray}\label{eq18}
&& f_r(x,v) =\rho_r \theta(-v) \, \theta \left(\frac{v^2}{2} -ax -\left(\frac{v_0^2}{2} -ax_0\right) \right)\qquad \\*\nonumber && \times  \sqrt{\frac{m}{2\pi k_BT}} \exp{\left\{-\frac{m}{k_BT} \left[\frac{v^2}{2} + a\left(\frac{L}{2} -x\right)\right]\right\}}
\end{eqnarray}
Using (\ref{eq5}) we find that in this case the contribution to the current due to ions from the right hand reservoir is:
\begin{eqnarray}\label{eq19}
j_r = -\rho_r \sqrt{\frac{k_BT}{2\pi m}} \exp{\left\{ -\frac{E_0}{k_BT} - \frac{ma}{k_BT}\left(\frac{L}{2}-x_0\right)\right\}}
\end{eqnarray}

(2) $E_0 < ma \left( x_0 + L/2\right)$. In this case, ions which pass through the barrier at $x_0$ do not necessarily have sufficient energy to overcome the remaining part of the field between $x_0$ and  $-L/2$. The only ions which contribute to the current are those entering the channel with velocity $v$, such that:
\begin{equation}\label{eq20}
\frac{mv^2}{2} > maL
\end{equation}
These make a contribution to the distribution function:
\begin{eqnarray}\label{eq21}
&& f_r(x,v) = \rho_r \theta(-v) \, \theta \left(\frac{v^2}{2} +a\left(\frac{L}{2}-x\right) - aL \right)\qquad \\*\nonumber && \times \sqrt{\frac{m}{2\pi k_BT}} \exp{\left\{-\frac{m}{k_BT} \left[\frac{v^2}{2} + a\left(\frac{L}{2} -x\right)\right]\right\}}
\end{eqnarray}
which results in a contribution to the current:
\begin{equation}\label{eq22}
j_r = -\rho_r  \sqrt{\frac{k_BT}{2\pi m}} \exp{\left\{ -\frac{maL}{k_BT} \right\}}
\end{equation}

The total current $j=j_l+j_r$ for model A is therefore:
\begin{eqnarray}\label{eq23}
 j && = \sqrt{\frac{k_BT}{2\pi m}}\bigg[\rho_l  -\rho_r  \theta \left(E_0 - ma \left(x_0 + \frac{L}{2}\right) \right) \\*\nonumber && \qquad \qquad \times \exp{\left\{ -\frac{E_0}{k_BT} - \frac{ma}{k_BT}\left(\frac{L}{2}-x_0\right)\right\}} \\*\nonumber && -\rho_r \theta \left(ma \left(x_0 + \frac{L}{2}\right)-E_0 \right) \exp{\left\{ -\frac{maL}{k_BT} \right\}}\bigg]
\end{eqnarray}
Expression (\ref{eq23}) is, of course, only valid for positive values of $a$. The equivalent expression when $a < 0$ can easily be shown to be:
\begin{eqnarray}\label{eq23a}
j_{a<0} && = \sqrt{\frac{k_BT}{2\pi m}}\bigg[-\rho_r +\rho_l  \theta \left(E_0 + ma \left(\frac{L}{2}-x_0\right) \right)\qquad \\*\nonumber && \qquad \qquad \times\exp{\left\{ -\frac{E_0}{k_BT} + \frac{ma}{k_BT}\left(\frac{L}{2}+x_0\right)\right\}} \\*\nonumber && +\rho_l \theta \left(-ma \left(\frac{L}{2}-x_0\right)-E_0 \right) \exp{\left\{ \frac{maL}{k_BT} \right\}}\bigg]
\end{eqnarray}

\begin{figure}
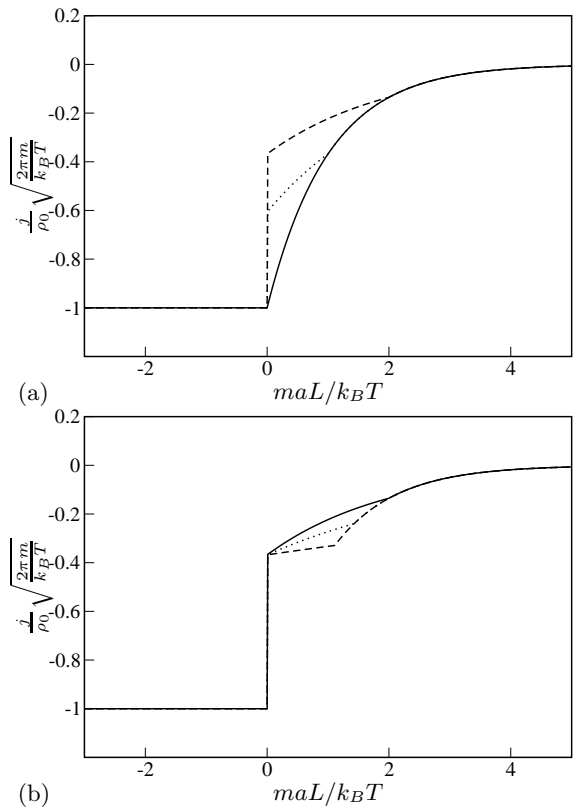

\begin{center}
\makebox[0pt][l]{(a)}{\rotatebox{0}{{\includegraphics[scale=0.3,clip=true]{fig3a.eps}}}}\hspace{0.5cm}\makebox[0pt][l]{(b)}{\rotatebox{0}{{\includegraphics[scale=0.3,clip=true]{fig3b.eps}}}}
\caption{Dimensionless current $\sqrt{2\pi m/(k_BT)}j/\rho_0$ for Model A (Equations (\ref{eq23}) and (\ref{eq23a})) as a function of $maL/(k_BT)$, ($\rho_0$ defined as in Figure \ref{fig2}), for $\rho_r=\rho_0$; $\rho_l=0$.  (a): Energetic trap fixed at centre of pore, $x_0/L=0$. Solid line: $E_0/(k_BT)=0$, dotted line: $E_0/(k_BT)=0.5$, dashed line: $E_0/(k_BT)=1$. (b): Height of barrier fixed, $E_0/(k_BT)=1$. Solid line: $x_0/L=0$, dotted line: $x_0/L = 0.2$, dashed line: $x_0/L=0.4$. \label{fig3}}
\end{center}
\end{figure}

The current-voltage curves for model A are shown in Figure \ref{fig3}, in dimensionless form as in Figure \ref{fig2}. For clarity, we consider the case where only the right-hand reservoir contains ions: $\rho_l=0$. In Figure \ref{fig3}a, the position of the trap is fixed at $x_0=0$ and the barrier height $E_0/k_BT$ is increased. When $maL/k_BT$ exceeds the critical value, given by $maL=E_0/(1/2+x_0/L)$, $j$ no longer depends on $E_0$ and all the curves become identical. However, in the regime where $maL<E_0/(1/2+x_0/L)$, $j$ depends strongly on $E_0$, being increased on increasing the barrier height. This can be easily understood, since ions which come from the right-hand reservoir and are impeded by the barrier make a negative contribution to the current. An interesting general observation can be made here, that the presence of an energetic barrier can have the effect of increasing the ionic current. In Figure \ref{fig3}b, the barrier height is fixed ($E_0/k_BT=1$) and the trap is moved towards the right-hand end of the pore. The current-voltage characteristics are seen to be rather sensitive to the position of the trap in the regime $maL<E_0/(1/2+x_0/L)$, although there is no dependence for larger $maL/k_BT$. As $x_0$ increases, $j$ decreases, although the value as $a \to 0$ remains unchanged.

We also consider an alternative energetic barrier model, model B. Here, ions encountering the trap with energy greater than the barrier height, $mv^2/2 > E_0$, do not continue unperturbed, as in model A, but instead  lose energy $E_0$, being released by the trap with reduced velocity $v'$, where $|v'| = \sqrt{v^2-v_0^2}$. Less energetic ions with  $ m v^2/2 < E_0$ are stopped by the trap, as in model A. Following a line of reasoning as for model A, one finds that the total current (when $a > 0$) is:
\begin{equation}\label{eq24}
j = \sqrt{\frac{k_BT}{2\pi m}}\left[\rho_l-\rho_r \exp{\left\{ -\frac{m}{k_BT}\left(aL+\frac{E_0}{m}\right) \right\}}\right]
\end{equation}
Note that for model B, the current $j$ does not depend on the position $x_0$ of the trap. Comparing expression (\ref{eq24}) with (\ref{eq15}) and (\ref{eq10}), we see that the current in model B with energy barrier $E_0$ is identical to that  through the channel with stopping traps investigated at the beginning of this section, if $E_0/k_BT = \int_{-L/2}^{L/2}\rho(x')\, dx'$.

\section{Finite channel with friction}\label{sec4}
We now turn to a model where no stopping traps are present  ($\rho(x)=0$), but ions undergo frictional collisions inside the pore (of finite length $L$).  The thermalizing effect on the ion of these  collisions with the channel surface and with other molecules (e.g. water) is modelled by a Langevin mechanism, represented by a Fokker-Planck operator, so that the stationary state kinetic equation (\ref{eq1}) now becomes: 
\begin{equation}\label{eq25}
\left(v\frac{\partial}{\partial x} + a \frac{\partial}{\partial v} \right) f(x,v) = \gamma \frac{\partial}{\partial v}\left(v+\frac{k_BT}{m}\frac{\partial}{\partial v}\right)f(x,v)
\end{equation}
Equation (\ref{eq25}) may be solved subject to boundary conditions specifying the incoming particle fluxes from the left ($x=-L/2$) and from the right ($x=L/2$), {\em{i.e.}}:
\begin{subequations}\label{eq26}
\begin{equation}
\int_0^\infty v f\left(-\frac{L}{2},v\right)dv = \rho_l \sqrt{\frac{k_BT}{2\pi m}}
\end{equation}
\begin{equation}
\int_{-\infty}^0 v f\left(\frac{L}{2},v\right)dv = -\rho_r \sqrt{\frac{k_BT}{2\pi m}}
\end{equation}
\end{subequations}

In the limit of an infinitely long channel ($L\to \infty$), the distribution must be homogeneous ($f(x,v) \to f(v)$); the solution of the corresponding Fokker-Planck equation ({\em{i.e.}} (\ref{eq25}) without the $v \partial / \partial x$ operator in the free flow term) is:
\begin{equation}\label{eq28}
f(v) \sim \exp{\left\{-\frac{m}{2k_BT}\left(v-\frac{a}{\gamma}\right)^2\right\}}\end{equation}
On the other hand, a particular inhomogeneous solution of Equation (\ref{eq25}) in a finite channel is:
\begin{equation}\label{eq29}
f(x,v) \sim \exp{\left\{-\frac{m}{k_BT}\left(\frac{v^2}{2}-ax\right)\right\}}\end{equation}
We now look for a solution of the Fokker-Planck equation (\ref{eq25}) for finite $L$, satisfying the boundary conditions (\ref{eq26}), in the form of a linear combination of the two solutions (\ref{eq28}) and (\ref{eq29}):
\begin{eqnarray}\label{eq30}
\nonumber f(x,v) &=& \sqrt{\frac{m}{2\pi k_BT}}  \Bigg [ A \exp{\left\{   -\frac{m}{k_BT}\left(\frac{v^2}{2}-ax\right)\right\}} \\* && + B\exp{\left\{-\frac{m}{2k_BT}\left(v-\frac{a}{\gamma}\right)^2\right\}}\Bigg]
\end{eqnarray}
Distribution (\ref{eq30}) indeed satisfies (\ref{eq25}) for all values of the coefficients $A$ and $B$. Imposing the boundary conditions (\ref{eq26}), we obtain:
\begin{equation}\label{eq31}
B=\frac{\rho_l \, \exp{\left\{\frac{maL}{2k_BT}\right\}}-\rho_r\,\exp{\left\{-\frac{maL}{2k_BT}\right\}}}{X}
\end{equation}
where
\begin{eqnarray}\label{eq31a}
&&X=2\sinh{\left(\frac{maL}{2k_BT}\right)}\,\exp{\left\{-\frac{ma^2}{2k_BT\gamma^2}\right\}}\\*\nonumber && + \frac{a}{\gamma}\sqrt{\frac{2\pi m}{k_BT}}\left[\cosh{\left(\frac{maL}{2k_BT}\right)}+ \sinh{\left(\frac{maL}{2k_BT}\right)}\int_{-\frac{a}{\gamma}}^{\frac{a}{\gamma}}\phi^T(v)\, dv\right]
\end{eqnarray}

and
\begin{equation}\label{eq32}
A=\frac{1}{2\sinh{\left(\frac{maL}{2k_BT}\right)}}\left[\rho_r-\rho_l + \frac{a}{\gamma}\sqrt{\frac{2\pi m}{k_BT}}B\right]
\end{equation}
The current through the channel can then be calculated using Equation  (\ref{eq5}):
\begin{equation}\label{eq33}
j=\frac{a}{\gamma}B
\end{equation}
Expression (\ref{eq33}) for the current simplifies greatly in the limit  of vanishing applied field ($a \to 0$), when the only driving force is diffusion under the action of the density gradient $(\rho_l-\rho_r)/L$. Substituting (\ref{eq31}) into (\ref{eq33}), one finds:
\begin{equation}\label{eq34}
\lim_{a\to 0} j = \sqrt{\frac{k_BT}{2 \pi m}}\frac{\rho_l-\rho_r}{\left(1+\gamma L \phi^T(0)\right)}
\end{equation}
{\em{i.e.}} the friction reduces the current by a factor $1/\left(1+\gamma L \phi^T(0)\right)$ compared to the free flow result (\ref{eq7}).  The $a \to 0$ limit of the distribution function $f(x,v)_{a \to 0}$ and the resulting density profile are discussed in Appendix \ref{appb}: $f(x,v)_{a \to 0}$ for a given $L$ and $\gamma$ turns out not to be everywhere positive,  pointing to a fundamental difficulty in applying the Fokker-Planck equation (\ref{eq25}) in a system of finite spatial extension $L$. 

Another interesting case is the limit of strong friction ($\gamma \to \infty$). $B$ then becomes:
\begin{equation}\label{eq35}
\lim_{\gamma \to \infty} B = \frac{\rho_l - \rho_r \exp{\left\{-\frac{maL}{k_BT} \right\}}}{1-\exp{\left\{-\frac{maL}{k_BT} \right\}}}
\end{equation}
and the resulting current, given by inserting (\ref{eq35}) in (\ref{eq33}), reduces to the classic GHK expression \cite{goldman,hodgkin}, which arises from solving the 1-d diffusion equation in a constant external field \cite{hille}. 

GHK theory predicts that the current increases linearly with voltage across the channel for large voltages. However, the behaviour of $j$ for large $a$ in this model is considerably different: fixing $\gamma$ and taking the limit $a \to \infty$ in Equation (\ref{eq31}), we find that the current saturates for large applied fields:
\begin{subequations}\label{eq36}
\begin{eqnarray}
\lim_{a \to \infty} j &=& \rho_l \sqrt{\frac{k_BT}{2\pi m}}\label{equationa}
\\
\lim_{a \to -\infty} j &=& -\rho_r \sqrt{\frac{k_BT}{2\pi m}}\label{equationa}
\end{eqnarray}
\end{subequations} 
Equations (\ref{eq36}), which are identical to the saturation values of the current for the models presented in section \ref{sec3}, correspond to the situation where all ions crossing the channel in the direction of the field contribute to the current and all ions attempting to penetrate the channel against the field are turned back.

An important experimental quantity is the ``reversal potential'': the voltage across the channel for which the total ionic current is zero. In the case where the ionic species are the same in the two reservoirs, the acceleration  $a_0$ at which the current is zero is given by cancelling the numerator of (\ref{eq31}), which yields:
\begin{equation}\label{eq38}
a_0 = \frac{k_BT}{mL}\ln{\left[\frac{\rho_r}{\rho_l}\right]}
\end{equation}
Expression (\ref{eq38}) is identical to the GHK prediction. However, if the reservoirs contain different species: for example potassium on one side and sodium on the other, the model will no longer agree with GHK theory. 

\begin{figure}
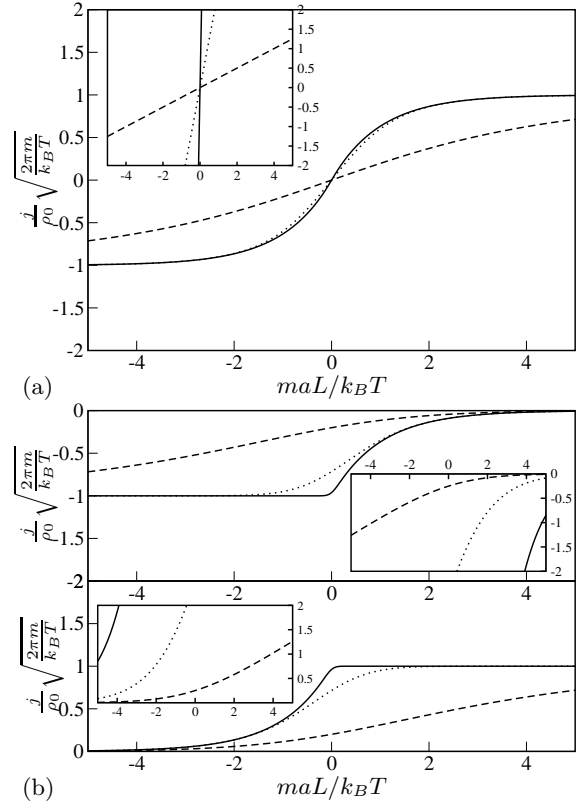

\begin{center}
\makebox[0pt][l]{(a)}{\rotatebox{0}{{\includegraphics[scale=0.3,clip=true]{fig4a.eps}}}}\hspace{0.5cm}\makebox[0pt][l]{(b)}{\rotatebox{0}{{\includegraphics[scale=0.3,clip=true]{fig4b.eps}}}}
\caption{Dimensionless current $\sqrt{2\pi m/(k_BT)}j/\rho_0$ as a function of $maL/(k_BT)$, for values of dimensionless friction $L\gamma \sqrt{m/(k_BT)}$ of 0.1 (solid lines), 1.0 (dotted lines) and 10.0 (dashed lines). (a): Equal reservoir densities $\rho_r=\rho_l=\rho_0$; inset shows results predicted by GHK theory. (b) Reservoir densities $\rho_r=\rho_0$; $\rho_l=0$ (above) and $\rho_r=0$; $\rho_l=\rho_0$ (below); again, inset shows results of GHK theory.\label{fig4}}
\end{center}
\end{figure}

Plots of the dimensionless current $\sqrt{2\pi m/(k_BT)}j/\rho_0$ versus $maL/(k_BT)$  are shown in Figure \ref{fig4}, for values of the dimensionless combination $L\gamma \sqrt{m/(k_BT)}$ of $0.1$, $1.0$ and $10.0$. In both Figure \ref{fig4}a and Figure \ref{fig4}b, the insets show the results of GHK theory. Figure \ref{fig4}a shows the current through the channel when the ionic concentrations in the two reservoir are equal ($\rho_r=\rho_l$). While GHK theory predicts linear asymptotic behaviour, the current given by Equation (\ref{eq33}) shows saturation as $|a| \to \infty$. As the friction coefficient decreases, the current-voltage relation becomes steeper and deviates further from the GHK results. In Figure \ref{fig4}b, the current shown in  Figure \ref{fig4}a is divided into the contributions of ions originating in the right (shown above) and left (shown below) reservoirs. As expected, for large positive $a$, the current is due only to ions from the left, and for large negative $a$, it consists only of ions from the right.

\section{Infinite channel with friction and traps}\label{sec5}
We next address the full version of the model system described in section \ref{sec2}: single ions moving under the influence of constant accelerating field, a Fokker-Planck thermalizing mechanism and stopping traps. We shall consider only the case where the velocity of an ion encountering a trap is set to zero, irrespective of its initial velocity, and where the average distribution of the traps is uniform ($\rho(x)=\rho$). An analytic solution of the kinetic equation (\ref{eq1}) is presented in the limit of an infinitely long pore ($L \to \infty$). This solution may prove useful in analysing ion flow through carbon nanotubes or the long pores found in zeolites.

We first introduce dimensionless position and velocity variables $y$ and $u$:
\begin{subequations}\label{eq39}
\begin{eqnarray}
x&=&\frac{1}{\rho}\,y\label{equationa}
\\*
v&=&\sqrt{\frac{k_BT}{m}}u\label{equationb}
\end{eqnarray}
\end{subequations}
as well as a dimensionless distribution function $F(y,u)$ through the transformation 
\begin{eqnarray}\label{eq41}
&&\nonumber f(x,v) dxdv = f\left(\frac{1}{\rho}y, \sqrt{\frac{k_BT}{m}}u\right)\frac{1}{\rho}\sqrt{\frac{k_BT}{m}}\, dy \, du\\* &&\qquad \qquad \equiv  F(y,u)\, dy\, du
\end{eqnarray}
Using (\ref{eq39}) and (\ref{eq41}), the kinetic equation (\ref{eq1}) becomes:
\begin{eqnarray}\label{eq42}
&&\left(\beta u \frac{\partial}{\partial y} + \alpha \frac{\partial}{\partial u}\right)F(y,u) \\*\nonumber &&= \beta \left\{ \delta(u) \int_{-\infty}^{+\infty} dw |w| F(y,w) - |u| F(y,u)\right\}\\*\nonumber && \qquad + \frac{\partial}{\partial u}\left(u + \frac{\partial}{\partial u}\right)F(y,u)
\end{eqnarray}
where the dimensionless coefficients $\alpha$ and $\beta$ are defined by:
\begin{subequations}\label{eq43}
\begin{eqnarray}
\alpha&=&\frac{a}{\gamma}\sqrt{\frac{m}{k_BT}}\label{equationa}
\\*
\beta&=&\frac{\rho}{\gamma}\sqrt{\frac{k_BT}{m}}\label{equationb}
\end{eqnarray}
\end{subequations}
We were unable to solve the inhomogeneous equation (\ref{eq42})  analytically. 
 An analytic solution may, however, be obtained in the in the limit of an infinitely long pore ($L \to \infty$), when the ion distribution no longer depends on $y$ and the problem is spatially  homogeneous.  Equation (\ref{eq42}) then simplifies to:
\begin{eqnarray}\label{eq45}
 \nonumber \alpha \frac{d F(u)}{d u} &=& \beta \left\{ \delta(u) \int_{-\infty}^{+\infty} dw |w| F(w) - u F(u)\right\} \\* && + \frac{d}{d u}\left(u + \frac{d}{d u}\right)F(u)
\end{eqnarray}
The solution of Equation (\ref{eq45}) is obtained as sketched in Appendix \ref{appc}. The result is:
\begin{eqnarray}\label{eq50} 
F(u) &=& A\exp{\left[-\frac{(u-\alpha)^2}{4}\right]}\times \\*\nonumber && \bigg\{\theta(u)\,D_{\beta(\beta+\alpha)}(\alpha{+}2\beta)\,D_{\beta(\beta-\alpha)}(u{-}\alpha{+}2\beta) \\\nonumber && + \theta(-u)\,D_{\beta(\beta-\alpha)}({-}\alpha{+}2\beta)\,D_{\beta(\beta+\alpha)}({-}u{+}\alpha{+}2\beta)\bigg\}
\end{eqnarray}
where the $D_p(z)$ are parabolic cylinder functions. The constant $A$ determines the number density of ions inside the channel (in a finite channel this would be set by the reservoir densities): here, we assume one ion per unit channel length, so that $A$ can be obtained numerically from the normalization condition $\int_{-\infty}^{+\infty} F(u) du = 1$.

In the limit $\beta \to 0$, {\em{i.e.}} in the absence of traps, (\ref{eq50}) reduces to the result (\ref{eq28}) (noting that $D_0(z) = \exp{\left[-z^2/4\right]}$ and reverting to dimensional units), and the current is linear in  $\alpha$. In the limit $\alpha \to 0$, {\em{i.e.}} in the absence of acceleration, the solution (\ref{eq50}) is seen to be an even function of $u$, so that the current $j$ vanishes, as expected for an infinitely long, spatially homogeneous channel.

\begin{figure}
\begin{center}
{\rotatebox{0}{{\includegraphics[scale=0.3,clip=true]{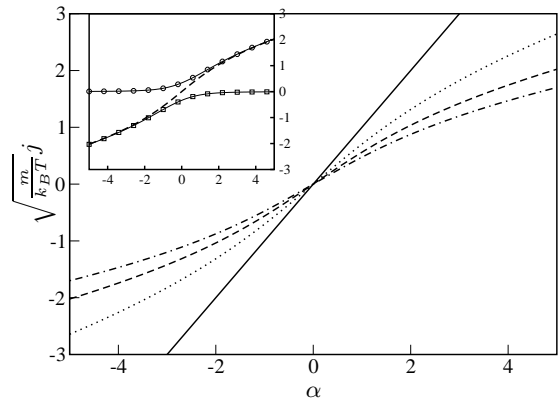}}}}
\caption{Current $J=\sqrt{m/(k_BT)}j$ as a function of $\alpha$, for values of $\beta$ of 0.0 (solid line), 0.25 (dotted line), 0.5 (dashed line) and 0.75 (dot-dashed line). Inset: Total $J$ (dashed line) as well as components of $J$ towards the right (circles) and towards the left (squares), for the case where $\beta=0.5$.\label{fig5}}
\end{center}
\end{figure}

In the general case, when both $\alpha$ and $\beta \ne 0$, the current must be calculated by numerical integration, after substituting (\ref{eq50}) in (\ref{eq5}). Figure \ref{fig5} shows the current $J=\int_{-\infty}^{\infty}u F(u) du = \sqrt{m/(k_BT)}j$, as a function of $\alpha$ for $\beta$ between $0$ and $0.75$. The inset shows the forward and backward components of $J$ when $\beta=0.5$ (given by integrating over the coefficients of $\theta(u)$ and $\theta(-u)$ in Equation (\ref{eq50})). There is a qualitative difference in the behaviour of the current when $\beta=0$, where the relation between $J$ and $\alpha$ is linear, as noted above, and when $\beta \ne 0$, where it is non-linear. Thus even a very small density of stopping traps (for example, due to defects or impurities) can have a dramatic effect on the current flowing through the pore. On estimating typical values of the  physical parameters  $a$, $\gamma$, $m$ and $\rho$, we find that $\alpha$ and $\beta$ are in fact likely to be small, perhaps of order $0.01-0.1$.

\section{Finite channel with thermalizing centres and traps}\label{sec6}
In Sections \ref{sec4} and \ref{sec5}, the effect of friction and thermalization on the motion of single ions was modelled by the Fokker-Planck collision operator. For the channel of finite length, this leads to the fundamental problem that  imposing the incoming ion fluxes from the reservoirs at both ends of the channel results in a stationary distribution function $f(x,v)$ which is not positive definite (see Appendix \ref{appb}). In this section, we therefore replace the Fokker-Planck mechanism by  an alternative thermalization process. We consider a model in which the 1-d channel contains $N$ ``thermalizing centres'', at positions $x_i$, such that:
\begin{equation}
-\frac{L}{2} < x_{1} < x_{2} < ... < x_{N} < \frac{L}{2}     \label{6.1}
\end{equation}
When an ion reaches a thermalizing centre, its incoming velocity $v$ is replaced by a new velocity $v'$ drawn from a Maxwell distribution $\phi^T(v')$, Equation (\ref{eq3}). The channel also contains a series of ``energy barriers'', of the type denoted ``Model A'' in section \ref{sec3}: a barrier of height $E_i$ temporarily stops an ion with kinetic energy $m v^2/2 < E_i$ but has no effect if $m v^2/2 > E_i$. An energy barrier of height $E_i$ is located between each pair of neighbouring thermalizing centres  at $x_{i-1}$ and $x_i$: if $E_i=0$, this is equivalent to having no energy barrier present. In between encounters with thermalizing centres and energy barriers, ions move with constant acceleration $a$, which is taken to act towards the right.

We now analyze the stochastic process defined by this model, leading to an exact calculation of the stationary current $j$. The key quantity is the probability $p(i \to i+1)$ that an ion which is thermalized by the centre at $x_i$, subsequently encounters the next thermalizing centre at $x_{i+1}$ ({\em{i.e.}} it reaches $x_{i+1}$ before $x_{i-1}$). We first note that an ion which leaves the centre at $x_i$, moving towards the left, requires a minimal energy $e(i,i-1)$ to penetrate the field and energetic barrier and reach the centre at $x_{i-1}$, where:
\begin{equation}
e(i,i-1) = ma(x_{i}-x_{i-1}) + E_{i} \label{6.2}
\end{equation}
We shall adopt the convenient notation $x_{0}=-L/2$ and $x_{N+1}=L/2$, so that Equation (\ref{6.2}) remains valid for $i=1$ and for $i=N+1$. Ions leaving $x_i$ towards the left with energy less than $e(i,i-1)$ will be stopped and re-accelerated towards the right, returning to  $x_i$. Since ions leave the thermalizing centre with a Maxwell velocity distribution, the probability  $w(i, i-1)$, that an ion leaving $x_i$ (in either direction) has energy less than $e(i,i-1)$ is given by:
\begin{equation}
w(i,i-1) = \frac{2}{\sqrt{\pi}}\int_{0}^{\sqrt{e(i,i-1) / k_{B}T}}du\; {\rm exp}(-u^{2}) \label{6.3}
\end{equation}
An ion that leaves $x_i$ towards the right, on the other hand, will certainly reach the thermalizing centre at $x_{i+1}$, regardless of whether it is stopped by the energetic barrier $E_{i+1}$. Thus on leaving $x_i$, an ion may be sent to the right (with probability $1/2$), and reach $x_{i+1}$, or it may be sent to the left, be stopped and return to $x_i$ (with probability $w(i,i-1)/2$), or lastly it may be sent to the left and reach $x_{i-1}$. 

The probability $p(i \to i+1)$ that an ion leaving $x_i$ reaches $x_{i+1}$ before $x_{i-1}$ can be found by summing over all the possible ways that this might happen.  The $n$-th term in the series corresponds to the scenario where an ion is sent $n$ times to the left ({\em{i.e.}} towards $x_{i-1}$) and returns to $x_i$ before eventually being sent to the right ({\em{i.e.}} towards $x_{i+1}$). We thus obtain a geometric series:
\begin{eqnarray}\label{6.4}
\nonumber p(i\to i+1)&=&\frac{1}{2}+\frac{1}{2}\left[\frac{w(i,i-1)}{2} \right]+\frac{1}{2}\left[\frac{w(i,i-1)}{2} \right]^{2}+... \\* & 
 =& \frac{1}{2-w(i,i-1)}
\end{eqnarray}
The probability for the transition in the opposite direction ($x_i$ to $x_{i-1}$) is then clearly:
\begin{equation} \label{6.5}
p(i\to i-1)= 1- p(i\to i+1) = \frac{1-w(i,i-1)}{2-w(i,i-1)}                     \end{equation}
The probabilities (\ref{6.4}) and (\ref{6.5}) may now be used to determine the probability $q_i$ that an ion starting from $x_i$ eventually leaves the channel through the right end at $x_{N+1}=L/2$. The $q_i$ satisfy the (detailed balance) equations:
\begin{equation}
q_{i}=p(i\to i-1)q_{i-1}+ p(i\to i+1)q_{i+1}  \label{6.6}
\end{equation}
with the boundary conditions
\begin{equation}\label{6.7}
q_{0}=0, \;\; q_{N+1}=1
\end{equation}
Defining the differences $\Delta_{i}=q_{i}-q_{i-1}$, one finds from (\ref{6.6}) that:
\begin{equation}
\frac{\Delta_{i+1}}{\Delta_{i}}=\frac{1}{p(i\to i+1)}-1=\frac{p(i\to i-1)}{p(i\to i+1)}  \label{6.8}
\end{equation}
Taking the product of both sides of Equation (\ref{6.8}) over $1 \le i \le n$ leads to:
\begin{equation}
\prod^{n}_{i=1}\frac{\Delta_{i+1}}{\Delta_{i}}=\frac{q_{n+1}-q_{n}}{q_{1}}=\prod^{n}_{i=1}\frac{p(i\to i-1)}{p(i\to i+1)}
\label{6.9}
\end{equation}
Summing both sides of the second equality in Equation (\ref{6.9}) over $1 \le n \le N$, we arrive at:
\begin{equation}
\frac{1-q_{1}}{q_{1}}=\sum_{n=1}^{N}\prod^{n}_{i=1}\frac{p(i\to i-1)}{p(i\to i+1)} \label{6.10}
\end{equation}
The only ions which make a contribution to the current are those which come from the left reservoir, pass through the whole channel and exit at the right end, and those which come from the right reservoir and exit at the left end. We now  calculate the probability $p(-L/2 \to L/2)$ that an ion entering the channel from the left reservoir will exit through the right end, and thus contribute to the current. On entering the pore at $-L/2$, the ion will reach the thermalizing centre at $x_1$ with probability $1$, so that the definition of $q_1$, together with Equations (\ref{6.10}), (\ref{6.4}) and (\ref{6.5}), lead to:
\begin{eqnarray}
 p(-L/2\to L/2)& = q_{1}&=\left\{  1 + \sum_{n=1}^{N}\prod^{n}_{i=1}\frac{p(i\to i-1)}{p(i\to i+1)} \right\}^{-1}\\*\nonumber && = \left\{  1 + \sum_{n=1}^{N}\prod^{n}_{i=1}[1-w(i,i-1)] \right\}^{-1}
\label{6.11}
\end{eqnarray}
Consider next an ion entering the channel from the right. It will reach the thermalizing centre at $x_N$ with probability  ${\rm exp}[-e(N+1,N)/k_{B}T]$. Therefore:
\begin{equation}
p(L/2\to -L/2) = (1-q_{N}){\rm exp}[-e(N+1,N)/k_{B}T]    \label{6.12}
\end{equation}
We find $q_N$ by setting  $n=N$ in Equation (\ref{6.9}) and using Equations (\ref{6.5}) and (\ref{6.12}):
\begin{eqnarray}\label{6.13}
1-q_{N}&=&q_{1}\prod^{N}_{i=1}\frac{p(i\to i-1)}{p(i\to i+1)}\\*\nonumber &=&p(-L/2\to L/2) \prod^{N}_{i=1}[1-w(i,i-1)] 
\end{eqnarray}
We now combine Equations (\ref{6.11}), (\ref{6.12}) and (\ref{6.13}) and  conclude that ions coming from the right reservoir contribute to the current with probability:
\begin{equation}\label{6.14}
p(L/2\to -L/2) =
 \frac{\prod^{N}_{i=1}[1-w(i,i-1)] \exp{\left\{\frac{-e(N+1,N)}{k_{B}T}\right\}}}{ 1 + \sum_{n=1}^{N}\prod^{n}_{i=1}[1-w(i,i-1)]} 
\end{equation}
The stationary current is given by the sum of  the incoming fluxes from the left and right reservoirs (found by assuming incoming Maxwell distributions), multiplied by the probabilities $p(-L/2\to L/2)$ and $p(L/2\to -L/2)$, which determine the extent to which the incoming flux is reduced by the action of the thermalizing  centers:
\begin{eqnarray}\label{6.15}
&& j = j_{l}+j_{r}\\*\nonumber  &&= \sqrt{\frac{k_{B}T}{2\pi m}}\left[\rho_l p(-L/2\to L/2) - \rho_r p(L/2\to -L/2)\right]                  
\end{eqnarray}
Inserting Equations (\ref{6.11}) and (\ref{6.14}) into (\ref{6.15}), we find:
\begin{eqnarray}
&&j =  \sqrt{\frac{k_{B}T}{2\pi m}}\times \\*\nonumber && \left[\frac{\rho_{l}-\rho_{r}\prod^{N}_{i=1}[1-w(i,i-1)] {\rm exp}[-e(N+1,N)/k_{B}T]}
{  1 + \sum_{n=1}^{N}\prod^{n}_{i=1}[1-w(i,i-1)]}\right]  \label{6.16}
\end{eqnarray}
Combined with formulae (\ref{6.2}) and (\ref{6.3}), Equation (\ref{6.16}) provides an explicit expression for $j$ as a function of parameters defining the internal structure of the channel.

In the absence of applied field ($a=0$) and energy barriers (all $E_i=0$), $w(i,i-1)=0$ and the current (\ref{6.16}), now due to the effect of the thermalizing centres only, takes the particularly simple form:
\begin{equation}
j =\sqrt{\frac{k_{B}T}{2\pi m}}\left[\frac{ \rho_{l}-\rho_{r}}{N+1}\right]                 \label{6.17}
\end{equation}
{\em{i.e.}} both incoming fluxes are reduced by the same factor $1/{(N+1)}$. The prediction (\ref{eq34}) of the FP equation under the same conditions ($a=E_i=0$) coincides with (\ref{6.17}), provided:
\begin{equation}
\frac{N}{L}=\gamma \sqrt{m/2\pi k_{B}T} \label{6.18}
\end{equation}
The ``effective friction'' introduced by the thermalizing centres is thus proportional to their density. Physically, (\ref{6.18}) also means that the relaxation time $\gamma^{-1}$ is of the order of the time taken by an ion to cover the average distance $L/N$ between the thermalizing centres with velocity $\sqrt{k_{B}T/m}$. This equivalence between the FP and thermalizing centre results does not hold, however, in the presence of an accelerating field ($a \ne 0$).

As in the case described by the FP collision term (cf. Equation (\ref{eq36})), the current $j$ saturates for large applied fields ($a \to \infty$) at the value $j_l$. This is because all ions coming from the left are driven through the channel by the strong field, while no ions  are able to cross the channel successfully from the right.

Formula (\ref{6.16}) simplifies greatly when the thermalizing centres are evenly distributed and in the absence of energy barriers, {\em{i.e.}} when $(x_{i}-x_{i-1})=L/(N+1)$ and $E_i=0$ for all $1\le i \le N$. In that case:
\begin{equation}
j =  \frac{1-s}{1-s^{N}}\sqrt{\frac{k_{B}T}{2\pi m}}\left[ \rho_{l}-\rho_{r}s^{N}\exp{\left\{-\frac{maL}{(N+1)k_{B}T}\right\}}\right]               \label{6.19}
\end{equation}
where
\begin{equation}
s = \frac{2}{\sqrt{\pi}}\int^{\infty}_{\sqrt{\frac{amL}{(N+1)k_{B}T}}} du\, {\rm exp}(-u^{2})\label{6.20}
\end{equation}

\begin{figure}
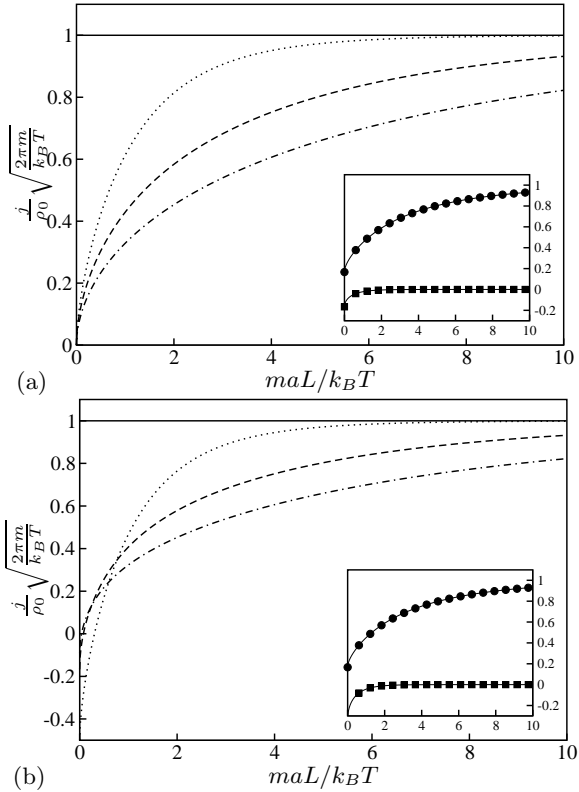

\begin{center}
\makebox[0pt][l]{(a)}{\rotatebox{0}{{\includegraphics[scale=0.3,clip=true]{fig6a.eps}}}}\hspace{0.5cm}\makebox[0pt][l]{(b)}{\rotatebox{0}{{\includegraphics[scale=0.3,clip=true]{fig6b.eps}}}}
\caption{Dimensionless current $\sqrt{2\pi m/(k_BT)}j/\rho_0$ as a function of $maL/(k_BT)$, where the reservoir densities are $\rho_l=C_l \rho_0$ and $\rho_r=C_r \rho_0$, for channels containing an increasing number $N$ of evenly spaced thermalizing centres. Energetic barrier heights $E_i$ are all set to zero.  Solid lines: $N=0$, dotted lines: $N=1$, dashed lines: $N=5$, dot-dashed lines: $N=10$. (a): Equal reservoir densities, $\rho_r=\rho_l$; $C_r=1$, $C_l=1$ (b): $\rho_r=2\rho_l$; $C_r=2$, $C_l=1$. The insets show the currents $j_l$ and $j_r$ (in dimensionless form) due to ions originating in the left (circles) and right (squares) reservoirs, when $N=5$.\label{fig6}}
\end{center}
\end{figure}

Figure \ref{fig6} shows the effect on the dimensionless current $\sqrt{2\pi m/(k_BT)}j/\rho_0$ of increasing the number $N$ of evenly spaced thermalizing centres, when there are no energetic barriers ($E_i=0$ for all $i$). In Figure \ref{fig6}a, the reservoir densities are equal, $\rho_l=\rho_r=\rho_0$. As $N$ increases, the current decreases, requiring larger values of $maL/(k_BT)$ to approach its asymptotic value. The inset shows the contributions to the (dimensionless) current from the left and right reservoirs when $N=5$. Figure \ref{fig6}b shows results for the same channel, when the density of ions in the right-hand reservoir is twice as large as that in the left: $\rho_r=2\rho_l=2\rho_0$. In this case, the negative contribution $j_r$ is increased and dominates for small values of $maL/(k_BT)$.

\begin{figure}
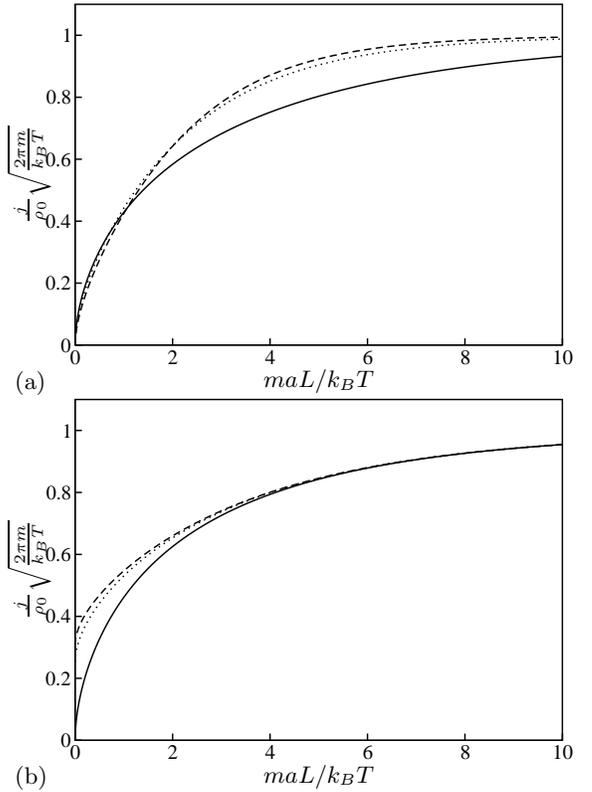

\begin{center}
\makebox[0pt][l]{(a)}{\rotatebox{0}{{\includegraphics[scale=0.3,clip=true]{fig7a.eps}}}}\hspace{0.5cm}\makebox[0pt][l]{(b)}{\rotatebox{0}{{\includegraphics[scale=0.3,clip=true]{fig7b.eps}}}}
\caption{Dimensionless current $\sqrt{2\pi m/(k_BT)}j/\rho_0$ as a function of $maL/(k_BT)$, for equal reservoir densities $\rho_l=\rho_r=\rho_0$. (a): Channel contains $N=5$ thermalizing centres, evenly spaced in the range $-b/2 \le x \le b/2$. All barrier heights $E_i$ are set to zero. Solid line: $b=L$, dotted line: $b=L/2$, dashed line: $b=L/4$. (b): Channel contains $N=4$ thermalizing centres, evenly spaced in the range $-L/2 \le x \le L/2$, and one energetic barrier $E_3 \ge  0$, located between the central pair of thermalizing centres. Solid line: $E_3/(k_BT)=0$, dotted line: $E_3/(k_BT)=1$, dashed line: $E_3/(k_BT)=2$.\label{fig7}}
\end{center}
\end{figure}

We have also investigated the effect of changing the spatial arrangement of the thermalizing centres, once again in the absence of energetic barriers ($E_i=0$ for all $i$).  In Figure \ref{fig7}a,  the channel contains $N=5$ thermalizing centres which are all located in the range $-b/2 \le x \le b/2$, where $b \le L$. Within this range the thermalizers are evenly spaced. Results are shown for equal reservoir densities, $\rho_l=\rho_r=\rho_0$. As $b$ decreases and the thermalizers become more localized in the middle of the pore, the current increases, approaching its asymptotic value for smaller values of $maL/(k_BT)$. However, the results for $b=0.5L$ (dotted lines) and $b=0.25L$ (dashed lines) are rather similar, indicating a limiting current-voltage relationship for small $b$. 

Interesting effects are obtained on including energetic barriers. Figure \ref{fig7}b shows results for a channel containing $N=4$ thermalizing centres, with a single barrier $E_i$, located between the second and third thermalizers ($i=3$), {\em{i.e.}} in the central one of the $5$ possible positions. Once again, $\rho_l=\rho_r=\rho_0$. As the barrier height $E_3$ is increased, the current increases, showing that inserting an impedance to ion passage can actually enhance the total ion flow through the channel. This apparently somewhat counter-intuitive result can in fact be easily understood. Let us consider an ion that is released by the thermalizing centre at $x_i$. If it is sent out towards the right, the ion will certainly reach $x_{i+1}$, while if it is sent to the left, it will be stopped and return to $x_i$ with probability $w(i, i-1)$, which is an increasing function of $E_i$. Thus increasing $E_i$ increases the chances of an ion eventually arriving at $x_{i+1}$, and thus enhances the current. This phenonenon may be of interest for biological ion channels, where the selectivity filter might play the role of an energetic barrier.

\section{Conclusion}
 In this paper, we have introduced some simple kinetic models for the transport of independent ions through narrow pores, under the influence of a constant acceleration, due to an applied external electric field. We consider only one-dimensional motion along the pore axis. Analytic results have been obtained for the stationary ion current $j$ and, in some cases, also for the stationary non-equilibrium distribution function $f(x,v)$. Boundary conditions at the ends of the pore are consistent with the presence of reservoirs containing ions at equilibrium, which determine the flux of ions entering the pore. The models include traps or energy barriers, which represent the temporary binding of an ion to polar residues lining the pore surface, and they also  account for the friction and thermalization due to collisions of the ions with molecules ({\em{e.g.}} water) inside the pore as well as with its inner, confining surface. Initially, this friction and thermalization was included using a Fokker-Planck operator in the kinetic equation. We were unable to find a solution for a pore of finite length $L$ when both traps and the Fokker-Planck description of friction are present. However, we have derived an  explicit solution for the homogeneous case of an infinitely long pore. For pores of finite length, solutions are given for models with stopping traps or energy barriers in the absence of friction. For stopping traps, the current does not depend on the spatial distribution of the traps, but for a single energy barrier, there may be a dependence on its position, depending on its mechanism of action. We have also solved the kinetic equation for the distribution function $f(x,v)$, for a finite channel in the  case where the Fokker-Planck mechanism is present but there are no traps or energy barriers. In this case, on imposing the incoming fluxes from the reservoirs at both ends of the pore,  $f(x,v)$ turns out not to be positive definite for short channels and/or small values of the friction constant $\gamma$. This unphysical behaviour
can be understood in terms of competing time scales for high velocity ions (which pass through the channel before they can be thermalized): the resulting ion current remains well-behaved, as does the number density profile. In view of this deficiency of the Fokker-Planck mechanism in a pore of finite length, we have introduced an alternative model, whereby ions are instead thermalized by encounters with a series of ``thermalizing centres'', located at given positions inside the pore. Energy barriers may also be present.

An important conclusion arising from all the models which were considered is that the current $j$ invariably saturates as a function of the external field (or equivalently the constant acceleration $a$), since it is limited by the incoming flux from the reservoirs.  This saturation behaviour contrasts with
the linear increase of j with voltage predicted by the classic GHK
result, which can be derived by solving the one-dimensional diffusion equation in the presence of a constant external field. A further interesting observation that emerges from this work is that the presence of stopping traps or energy barriers inside the pore can increase rather than decrease the ionic current. This is because the steady state flow of ions crossing the pore in the direction of the applied acceleration is unaffected by the traps or barriers, while the flow of ions against the field is reduced. When the former contribution is the dominant one, the current will be enhanced by the traps or barriers.

A question that arises is whether there is any correspondance between the two models of friction and thermalization considered in this work: the Fokker-Planck mechansim and the the model involving N thermalizing centres.  In the absence
of an applied field and of energy barriers, when ionic motion is driven only by the  concentration gradient across the pore, an  equivalence can be established between the two models (c.f. Equation (\ref{6.18})). However, under
more general conditions, we have found no one-to-one correspondance between the two descriptions of dissipation.

Future work will include a complete numerical analysis of the model involving both energy barriers and thermalizing centres.  We also
plan to extend the kinetic models to include the possibility of collective
ion permeation through a pore, by including ion-ion interactions. These are believed to play an important role in ion transport through some biological channels \cite{morais,zhou,berneche,berneche2}. Appropriate selection of parameters  such as the pore length $L$, the friction coefficient $\gamma$ or the number and positions of the thermalizing centres and the positions and heights of the energy barriers, to correspond to the structure of real ion channels, should allow the predictions of these kinetic models to be compared to measured current-voltage characteristics.

\appendix

\section{}\label{appa}
Here we derive an  analytic solution for the kinetic equation (\ref{eq9}), for the case where $\rho(x)=\rho$: {\em{i.e.}} we find the distribution function $f(x,v)$ for ions experiencing a uniform external electric field and traps distributed according to a Poisson law corresponding to a uniform average density $\rho$.

The distribution function $f(x,v)$ is split into the contributions $f^+(x,v)$ and $f^-(x,v)$ of ions moving to the right and to the left:
\begin{equation}\label{eqa1}
f(x,v) = \theta(v) f^+(x,v) + \theta(-v)f^-(x,v)
\end{equation}
Substituting (\ref{eqa1}) into (\ref{eq9}) (with $\rho(x)=\rho$), we obtain:
\begin{subequations}\label{eqa2}
\begin{eqnarray}
\left(v\frac{\partial}{\partial x} + a\frac{\partial}{\partial v} + \rho v\right)f^+(x,v) = 0 \label{equationa}
\\*
\left(v\frac{\partial}{\partial x} + a\frac{\partial}{\partial v} - \rho v\right)f^-(x,v) = 0\label{equationb}
\end{eqnarray}
\end{subequations}

Equations (\ref{eqa2}) imply that:
\begin{subequations}\label{eqa5}
\begin{eqnarray}
f^+(x,v) &=& \exp{[-\rho x]} F^+ \left(v^2/2 -ax\right)\label{equationa}
\\*
f^-(x,v) &=& \exp{[\rho x]} F^- \left(v^2/2 -ax\right)\label{equationb}
\end{eqnarray}
\end{subequations}
The as yet unknown functions $F^+$ and $F^-$ are linked by the requirement (\ref{eq6}) that the current $j$ be independent of $x$:
\begin{equation}\label{eqa4}
j=\int dv\, v f(x,v) = {\rm{constant}}
\end{equation}
Substituting (\ref{eqa5}) into (\ref{eqa4}), and defining $w=v^2/2-ax$, we obtain:
\begin{eqnarray}\label{eqa6a}
&&j={\rm{constant}} \\*\nonumber &&=  \exp{[-\rho x]}\int_{-ax}^{\infty}F^+(w)\,dw - \exp{[\rho x]}\int_{-ax}^{\infty}F^-(w)\,dw
\end{eqnarray}
Multiplying (\ref{eqa6a}) by $\exp{[\rho x]}$ and differentiating with respect to $x$, we obtain:
\begin{eqnarray}\label{eqa7}
&& a F^+(-ax) =  j\rho \exp{[\rho x]} \\*\nonumber && \qquad+ \left[2\rho \int_{-ax}^{\infty}F^-(w)\,dw + a F^-(-ax)\right]\exp{[2\rho x]} 
\end{eqnarray}
Equation (\ref{eqa7}) is valid inside the channel, {\em{i.e.}} for values of $x$ in the range $-L/2 \le x \le L/2$. The argument $-ax$ of $F^+$ and $F^-$ therefore ranges between $-aL/2$ and $aL/2$, so that the relationship (\ref{eqa7}) between $F^+(w)$ and $F^-(w)$ holds for $-La/2 \le w \le La/2$. Since $w=v^2/2-ax$, this corresponds to $v^2/2 \le a(L/2+x)$.

The distribution of particles moving against the field, $f^-(x,v)$, can be obtained using simple arguments. If a particle reaches position $x$ with negative velocity $v$, then it must have had energy $mu^2/2 = mv^2/2 + ma(L/2-x)$ at the right-hand channel entrance, and have encountered no traps over the distance $(L/2-x)$, since on encountering a trap an ion is re-accelerated by the field towards the right. Assuming a Maxwell distribution for the velocities of the incoming ions and noting that the Poisson probability for encoutering no traps over this distance is $\exp{[-\rho(L/2-x)]}$, we find that:
\begin{eqnarray}\label{eqa9}
&&f^-(x,v) = \exp{[\rho x]}\rho_r \sqrt{\frac{m}{2\pi k_BT}}\times \\*\nonumber &&\exp{\left[-\frac{m}{k_BT}\left(v^2/2-ax\right)\right]} \exp{\left[-\frac{L}{2}\left(\rho+\frac{ma}{k_BT}\right)\right]}
\end{eqnarray}
and thus from (\ref{eqa5}):
\begin{equation}\label{eqa9a}
F^-(w) = \rho_r \sqrt{\frac{m}{2\pi k_BT}}\exp{\left[-\frac{mw}{k_BT}\right]} \exp{\left[-\frac{L}{2}\left(\rho+\frac{ma}{k_BT}\right)\right]}
\end{equation}
Substituting the result (\ref{eqa9a}) into (\ref{eqa7}), we find that  for $v^2/2 \le a(L/2+x)$:
\begin{eqnarray}\label{eqa10}
\nonumber f^+(x,v) &&= j\frac{\rho}{a}\exp{\bigg[-\frac{\rho}{2a}v^2\bigg]} + \frac{\rho_r}{a}\bigg[\frac{m}{k_BT}+\frac{2\rho}{a}\bigg]\sqrt{\frac{k_BT}{2\pi m}}\\*\nonumber && \times \exp{\bigg[-\frac{m}{k_BT}\left(\frac{v^2}{2}-ax\right)\bigg]}\exp{\bigg[\rho x -\frac{\rho}{a}v^2\bigg]}\\* && \times \exp{\bigg[-\frac{L}{2}\left(\rho+\frac{ma}{k_BT}\right)\bigg]}
\end{eqnarray}
Using a similar argument to that above, we can determine the remaining part of $f^+(x,v)$, for $v^2/2 \ge a(L/2+x)$. Ions with $v^2/2 \ge a(L/2+x)$, moving towards the right, cannot have been stopped by a trap, since their energy is larger than that due to the accelerating field over the distance travelled in the pore. These ions must have entered the pore at $x=-L/2$ with energy $mu^2/2 = mv^2/2 - ma(L/2+x)$, and have encountered no traps over a distance $L/2+x$. Assuming a Maxwell distribution of incoming particles, we obtain a contribution to $f^+(x,v)$ of:
\begin{eqnarray}\label{eqa8}
&& \theta \left( {v^2}/{2}-a\left({L}/{2}+x\right)\right) \exp{[-\rho \left({L}/{2}+x\right)]}\times \\*\nonumber && \rho_l \sqrt{\frac{m}{2\pi k_BT}}\exp{\left[-\frac{m}{k_BT}\left({v^2}/{2}-a\left({L}/{2}+x\right)\right)\right]} 
\end{eqnarray}
Noting that at $x=-L/2$, $v^2/2 \ge a(L/2+x)$, we can use equations (\ref{eqa9}) and (\ref{eqa8}) for  $x=-L/2$ to calculate the current:
\begin{equation}\label{eqa12}
j=\sqrt{\frac{k_BT}{2\pi m}}\left\{\rho_l-\rho_r \exp{\left[-L\left(\rho+\frac{ma}{k_BT}\right)\right]}\right\}
\end{equation}
Substituting this result in (\ref{eqa10}), the final result for the distribution function $f(x,v)$ is:
\begin{eqnarray}\label{eqa11}
\nonumber &&f(x,v)= \theta(-v) \rho_r \phi^T(v)\exp{\left[\left(\frac{ma}{k_BT}+\rho\right)\left(x-L/2\right)\right]} \\* && \qquad \qquad+ \theta(v) \big\{ \theta \left( {v^2}/{2}-a\left({L}/{2}+x\right)\right) Y \\*\nonumber && \qquad\qquad  \qquad + \theta \left(a\left({L}/{2}+x\right)- {v^2}/{2}\right) Z\big\}
\end{eqnarray}
where 
\begin{subequations}
\begin{eqnarray}
&&Y=\rho_l\phi^T(v)\exp{\left[\left(\frac{ma}{k_BT}-\rho\right)\left(x+{L}/{2}\right)\right]} \label{equationa}
\\*
\nonumber&&Z=\sqrt{\frac{k_BT}{2\pi m}}\left(\rho_l-\rho_r \exp{\left[-L\left(\rho+\frac{ma}{k_BT}\right)\right]}\right)\\* && \qquad \qquad \times \frac{\rho}{a} \exp{\left[-\frac{\rho}{2a}v^2\right]} \\\nonumber &&\qquad + \left(\frac{2\rho k_BT}{am}+1\right) \exp{\left[-\frac{\rho}{a}v^2\right]}\\*\nonumber && \qquad \qquad \times \rho_r\phi^T(v) \exp{\left[\left(\frac{ma}{k_BT}+\rho\right)\left(x-{L}/{2}\right)\right]}
\end{eqnarray}
\end{subequations}
and $\phi^T(v)$ is the Maxwell distribution function, Equation (\ref{eq3}).
Equations (\ref{eqa12}) and (\ref{eqa11}) were derived for an accelerating field towards the right ($a > 0$). For $a<0$, the corresponding result for the current is:
\begin{equation}\label{jneg}
j=\sqrt{\frac{k_BT}{2\pi m}}\left\{\rho_l\exp{\left[-L\left(\rho-\frac{ma}{k_BT}\right)\right]}-\rho_r\right\}
\end{equation}

\section{}\label{appb}
In section \ref{sec4}, we constructed  the solution of the Fokker-Planck equation (\ref{eq25}), satisfying the boundary conditions given in Equation (\ref{eq26}). The solution consists of a linear combination of a non-equilibrium stationary homogeneous solution (\ref{eq28}), which gives rise to a constant current, and an inhomogeneous equilibrium state (\ref{eq29}), which does not contribute to the current. We thus found the stationary state (\ref{eq30}), which is of the form
\begin{equation}
f(x,v) = A \exp{ \left[ \frac{max}{k_BT}\right]}\phi^{T}(v) + B\phi^{T}\left(v-\frac{a}{\gamma}\right) \label{B1}
\end{equation}
where $\phi^{T}$ denotes the Maxwell distribution (\ref{eq3}). The corresponding current $j$ and density $n(x)$ of the ions are given by:
\begin{subequations}\label{B2}
\begin{eqnarray}
j&=&\frac{a}{\gamma}B  \label{equationa}
\\
n(x)&=&A \exp{ \left[ \frac{max}{k_BT}\right]} + B  \label{equationb} 
\end{eqnarray}
\end{subequations}
The imposed incoming ionic fluxes from the thermostats determine uniquely the values of $A$ and $B$ (Equations (\ref{eq31}) and (\ref{eq32})).

 When the ion density in the right thermostat is related to that in the left thermostat by the Boltzmann factor
\begin{equation}
   \rho_{r} = \rho_{l}\exp{ \left[ \frac{maL}{k_{B}T}\right]}    \label{B3}
\end{equation}
the coefficient $B$ vanishes and the distribution (\ref{B1}) reduces to the well known equilibrium solution of the FP equation (\ref{eq29}). When
 relation (\ref{B3}) does not hold, however,  equilibrium cannot occur and a stationary current will flow through the channel. It can be 
proved that in this situation the ionic number density $n(x)$ is  positive, but a difficulty arises when one considers the velocity distribution in the region of 
large velocities. In order to illustrate the problem let us study the simple limiting case of vanishing acceleration. From Equations (\ref{eq31}) and (\ref{eq32}) in the limit
of $a\to 0$ we get the asymptotic relations
\begin{equation}
A+B  = \frac{\rho_{l}+\rho_{r}}{2}  \label{B4}
\end{equation}
\begin{eqnarray}
(1+\gamma L\phi^{T}(0)) B & = & \gamma L\phi^{T}(0)\frac{\rho_{l}+\rho_{r}}{2}\\*\nonumber && + \frac{\gamma}{2\pi a\phi^{T}(0)}(\rho_{l}-\rho_{r})\label{B4a}
\end{eqnarray}
The stationary distribution (\ref{B1}) takes the form
\begin{eqnarray}\label{B5}
&&\lim_{a\to 0}f(x,v)  = \frac{\rho_{l}+\rho_{r}}{2}\phi^{T}(v)  \\*\nonumber && \qquad +  \frac{\gamma}{2\pi a\phi^{T}(0)}\frac{(\rho_{l}-\rho_{r})}{1+\gamma L\phi^{T}(0)}
\\*\nonumber &&\qquad\qquad \times \left( \phi^{T}(v-{a}/{\gamma})-\exp{ \left[ \frac{max}{k_BT}\right]}\phi^{T}(v) \right)  \nonumber
\end{eqnarray}
The evaluation of the point limit (at fixed values of the variable $v$) leads eventually to the distribution
\begin{equation}
f_{a=0}(x,v)= \left[ \frac{\rho_{l}+\rho_{r}}{2}+ \frac{(\rho_{l}-\rho_{r})\phi^{T}(0)}{1+\gamma L\phi^{T}(0)}
(v-\gamma x)\right] \phi^{T}(v)     \label{B6}
\end{equation}
The distribution (\ref{B6}) is an inhomogeneous solution of the the FP equation
\begin{equation}
v\frac{\partial}{\partial x} f(x,v) = \gamma \frac{\partial}{\partial v}\left(v+\frac{k_BT}{m}\frac{\partial}{\partial v}\right)f(x,v)  
\end{equation}
satisfying to the boundary conditions (\ref{eq26}). The current and the density profile, which is linear, are given by
\begin{subequations} 
\begin{eqnarray}
j & = & \frac{j_{l}-j_{r}}{1+ \gamma L\phi^{T}(0)} \label{equationa} 
 \\*
n(x) & = & \frac{\rho_{l}+\rho_{r}}{2}+ \frac{(\rho_{l}-\rho_{r})\phi^{T}(0)}{1+\gamma L\phi^{T}(0)}\gamma x  \label{equationb} 
\end{eqnarray}
\end{subequations}
Whereas $n(x)$ is positive everywhere within the channel, the complete distribution $f_{a=0}(x,v)$ is not positive definite. 
For example, when $\rho_{l}<\rho_{r}$, $f_{a=0}(x,v)$ turns negative for sufficiently large velocities and thus loses its physical 
meaning. Hence it seems that a physically acceptable inhomogeneous stationary state cannot be obtained from the FP equation.

       Notice that one can define a characteristic velocity $\gamma L$, associated with the finite length channel: ions with velocities larger than this will
cover the length of the channel in a time interval which is shorter than the FP thermalization time $\gamma^{-1}$.
It turns out that when $v<\gamma L$ the distribution function $f_{a=0}(x,v) > 0$, and so the difficulty appears only for velocities larger than $\gamma L$. Clearly,  when the friction coefficient $\gamma$ is large enough the unphysical region is 
quantitatively irrelevant  because of the negligibly small probability weight coming from  the Maxwell distribution. From a 
fundamental point of view, however, the solution (\ref{B6}) demonstrates that the FP equation is incompatible with the boundary conditions (\ref{eq26}) for a channel of finite length.  

\section{}\label{appc}
In this section, we describe the solution of the kinetic equation (\ref{eq45}), for ions flowing through a channel under the influence of a uniform accelerating field, a Fokker-Planck friction mechanism and stopping traps distributed on average uniformly, in the limit where the channel is infinitely long and the problem is spatially homogeneous.

The first term on the r.h.s. of Equation (\ref{eq45}) describes the effect of the stopping traps (in terms of a balance between loss of particles of dimensionless velocity $u$ and gain of those with $u=0$). This term may be  recast in the form:
\begin{eqnarray}\label{eq46}
&& \beta \left\{ \delta(u) \int_{-\infty}^{+\infty} dw |w| F(y,w) - |u| F(y,u)\right\}\\*\nonumber
&& = \frac{\beta}{2}\frac{d}{du} \Bigg\{ {\rm{sgn}}(u) \int_{-\infty}^{+\infty} dw \,|u-w| F(w) \\*\nonumber &&\qquad\qquad - |u| \int_{-\infty}^{+\infty} dw\,\, {\rm{sgn}}(u-w) F(w)\Bigg\}\\*\nonumber && = \frac{\beta}{2}\frac{d}{du} \left\{ -{\rm{sgn}}(u) \int_{-\infty}^{+\infty} dw \,w\, {\rm{sgn}}(u-w) F(w)\right\}
\end{eqnarray}
where ${\rm{sgn}}(u) = \theta(u) - \theta(-u)$.

Substituting (\ref{eq46}) into (\ref{eq45}), and integrating over $u$, one obtains the following relation:
\begin{eqnarray}\label{eq47}
&& \left(u-\alpha +\frac{d}{du} \right) F(u) \\*\nonumber && = \frac{1}{2}\beta {\rm{sgn}}(u) \int_{-\infty}^{+\infty} dw \,w\, {\rm{sgn}}(u-w) F(w) + C
\end{eqnarray}
The integration constant $C$ on the r.h.s. of this relation is determined by the boundary condition $\lim_{u \to \pm \infty} F(u) =0$, which shows that 
\begin{eqnarray}\label{eq48}
C=-\frac{\beta}{2}\int_{-\infty}^{+\infty} dw\, w F(w) = -\frac{\beta}{2}j
\end{eqnarray}
where $j$ is the dimensionless current. Gathering results, Equation (\ref{eq47}) may be cast in the form:
\begin{eqnarray}\label{eq49}
&& \left(u-\alpha +\frac{d}{du} \right) F(u) \\*\nonumber &&= -\beta \left\{ \theta(u) \int_{u}^{+\infty} dw \,w F(w) + \theta(-u) \int_{-\infty}^{u} dw \,w F(w)\right\} 
\end{eqnarray}

The structure of the integro-differential equation (\ref{eq49}) suggests seeking a solution of the form:
\begin{eqnarray}\label{eqb1}
F(u) = \theta(u) F^+(u) + \theta(-u) F^-(u)
\end{eqnarray}
where $F^+$ and $F^-$ satisfy the equations:
\begin{subequations}
\begin{equation}\label{eqb2}
\left(u-\alpha + \frac{d}{du}\right)F^+(u) = -\beta \int_u^\infty dw\, w F^+(w)
\end{equation}
\begin{equation}\label{eqb3}
\left(u-\alpha + \frac{d}{du}\right)F^-(u) = -\beta \int_{-\infty}^u dw\, w F^-(w)
\end{equation}
\end{subequations}
to be solved, subject to the boundary condition:
\begin{eqnarray}\label{eqb4}
F^+(0) - F^-(0) = 0
\end{eqnarray}
Differentiation of (\ref{eqb2}) leads to
\begin{eqnarray}\label{eqb5}
\nonumber  && \frac{d^2}{du^2} F^+(u) + (u-\alpha)\frac{dF^+(u)}{du} + (1-\beta u)F^+(u) \\*&& \qquad \qquad = 0
\end{eqnarray}
Seeking a solution of the form 
\begin{equation}
F^+(u) = \exp{\left[-(u-\alpha)^2/4\right]}G^+(u) 
\end{equation}
and substituting in (\ref{eqb5}), one arrives at the following differential equation for $G^+(u)$:
\begin{eqnarray}\label{eqb6}
\nonumber  && \frac{d^2 G^+(u)}{du^2} + \left[\frac{1}{2} + \beta(\beta-\alpha)  - \frac{(u-\alpha + 2\beta)^2}{4}  \right]G^+(u)\\*&& \qquad \qquad=0
\end{eqnarray}
the solution of which is:
\begin{eqnarray}\label{eqb7}
G^+(u) = D_{\beta (\beta-\alpha)}(u-\alpha+2\beta)
\end{eqnarray}
where $D_p(z)$ is a parabolic cylinder function. Hence:
\begin{eqnarray}\label{eqb8}
F^+(u) = \exp{\left[-{(u-\alpha)^2}/{4}\right]}D_{\beta (\beta-\alpha)}(u-\alpha+2\beta)\qquad
\end{eqnarray}
Proceeding along similar lines, one finds the solution of (\ref{eqb3}) in the form:
\begin{eqnarray}\label{eqb9}
F^-(u) = \exp{\left[-{(u-\alpha)^2}/{4}\right]}D_{\beta (\beta+\alpha)}(-u+\alpha+2\beta)\qquad
\end{eqnarray}
and hence:
\begin{eqnarray}\label{eqb10}
F(u) = &&  \exp{\left[-{(u-\alpha)^2}/{4}\right]}\\*\nonumber && \times \Bigg\{ A_1\, \theta(-u)\, D_{\beta (\beta+\alpha)}(-u+\alpha+2\beta) \\*\nonumber && \qquad + A_2 \, \theta(u)\, D_{\beta (\beta-\alpha)}(u-\alpha+2\beta) \Bigg\}
\end{eqnarray}
A relation between the coefficients $A_1$ and $A_2$ follows from the boundary condition (\ref{eqb4}), namely:
\begin{eqnarray}\label{eqb11}
A_1 D_{\beta (\beta+\alpha)}(\alpha+2\beta) = A_2 D_{\beta (\beta-\alpha)}(-\alpha+2\beta) 
\end{eqnarray}
from which the general solution (\ref{eq50}) in the main text follows.

\begin{acknowledgments}
The authors thank Bob Eisenberg for his advice. JP thanks JP-H for inviting him to Cambridge and EPSRC for financial support. RJA acknowledges the support of EPSRC and Unilever. 
\end{acknowledgments}


\end{document}